\documentclass[fleqn,10pt]{wlscirep}

\usepackage{amsmath}
\usepackage{amsfonts}
\usepackage{graphicx,color}
\usepackage[caption = false]{subfig}
\usepackage{float}

\newcommand{\ket}[1]{| #1 \rangle}

\title{Tuning quantum measurements to control chaos}

\author[1,*]{Jessica K. Eastman}
\author[2]{Joseph J. Hope}
\author[1,3]{Andr\'e R. R. Carvalho}

\affil[1]{Centre for Quantum Computation and Communication Technology, Department of Quantum Science, Research School of Physics and Engineering, The Australian National University, Canberra ACT 2601 Australia}
\affil[2]{Department of Quantum Science, Research School of Physics and Engineering, The Australian National University, Canberra ACT 2601 Australia}
\affil[3]{Centre for Quantum Dynamics, Griffith University, Nathan QLD 4111, Australia}

\affil[*]{jessica.eastman@anu.edu.au}

\begin{abstract}
Environment-induced decoherence has long been recognised as being of crucial importance in the study of chaos in quantum systems. In particular, the exact form and strength of the system-environment interaction play a major role in the quantum-to-classical transition of chaotic systems. In this work we focus on the effect of varying monitoring strategies, i.e. for a given decoherence model and a fixed environmental coupling, there is still freedom on how to monitor a quantum system. We show here that there is a region between the deep quantum regime and the classical limit where the choice of the monitoring parameter allows one to control the complex behaviour of the system, leading to either the emergence or suppression of chaos. Our work shows that this is a result from the interplay between quantum interference effects induced by the nonlinear dynamics and the effectiveness of the decoherence for different measurement schemes.  
\end{abstract}
\begin{document}

\flushbottom
\maketitle
%
%
\thispagestyle{empty}

\section*{Introduction}

Understanding how classical dynamics emerge from the more fundamental quantum theory has proven to be a subtle problem when the system in question exhibits chaos in the classical limit. Coherent interference effects lead to a rapid breakdown of the correspondence between the classical and quantum dynamics. The inclusion of decoherence effects destroys the interference and is a crucial step to achieve a smooth quantum to classical transition~\cite{Ott:1984,Dittrich:1987, Zurek:1994, Habib:1998, Pattanayak:2003, Carvalho:2004b}. 

Many studies of classically chaotic systems undergoing environmental coupling have focused on the ensemble average behaviour given by the master equation and the comparison of the classical phase space with its quantum counterpart via Wigner functions. Others have adopted an approach based on continuously monitored quantum systems~\cite{Dittrich:1990a,Spiller:1994,Brun:1996,Rigo:1996,Bhattacharya:2000, Ghose:2003, Habib:2006, Kapulkin:2008}. In this case, the monitoring is said to produce an ``unraveling" of the master equation in terms of individual stochastic quantum trajectories that evolve conditioned on the measurement record. Using this approach, it has been shown that the Poincar\'{e} section of a single quantum trajectory reproduces the corresponding classical strange attractors in the macroscopic limit~\cite{Spiller:1994, Brun:1996}, even when considering a few different monitoring strategies~\cite{Rigo:1996}. It also allowed a quantitative comparison between classical and quantum Lyapunov exponents as the effective size of the system varies~\cite{Bhattacharya:2000,Ota:2005,Habib:2006,Kapulkin:2008}. In general, when the classical motion is large compared to the quantum noise induced by the stochastic nature of the trajectories, the quantum Lyapunov exponent approaches the classical value~\cite{Bhattacharya:2000}, while there is a crossover to the quantum regime where noise predominates and chaos is suppressed~\cite{Ota:2005}. Interestingly, positive Lyapunov exponents have been found away from the classical limit~\cite{Habib:2006} but perhaps even more surprising is the fact that they have also been reported for parameters where the corresponding classical system is regular~\cite{Kapulkin:2008, Pokharel:2016}.

These results show not only that the onset of chaos at the quantum level is possible, but also that it has a rich behaviour due to the interplay between the strength of the nonlinear dynamics and the amount of noise introduced by the measurement back action. But quantum mechanics allows us to go beyond that and explore more complex scenarios where, even when the form and strength of the system-environment interaction are kept unchanged, different choices of measurement schemes can have a drastic effect on the dynamics of the system. This is the purpose of this contribution: we show that the Lyapunov exponent of the quantum system is sensitive to the choice of monitoring strategy and, consequently, one can control the degree of chaos in the system by tuning an easily accessible measurement parameter. Our results show that this effect originates from a fine balance between two competing factors: the appearance of interference at the quantum level due to the underlying classical nonlinear dynamics, and the effectiveness of certain monitoring schemes in destroying these very same interference fringes. 

Our starting point is the driven dissipative Duffing oscillator, a system that exhibits chaos for a wide range of parameters and that has been extensively studied both in the classical and quantum domains~\cite{Brun:1996,Ota:2005,Schack:1995}. Classically, the system is described by the equation of motion  
\begin{equation}
\ddot{x} + 2\Gamma\dot{x} + \beta^2x^3 - x = \frac{g}{\beta}\cos{(\Omega t)},
\label{eq:class}
\end{equation}
with dissipation $\Gamma$, driving amplitude $g$, and driving frequency $\Omega$. 
Quantum mechanically, the system dynamics is described by the master equation:
\begin{equation}
\label{eq:ME}
\dot{\rho} = -i[\hat H,\rho] + \left(\hat L \rho \hat L^\dagger - \frac{1}{2}\{\hat L^\dagger \hat L, \rho \}\right),
\end{equation}
where the term in parentheses represents the dissipative evolution in the Lindblad form~\cite{Lindblad:1976} with the operator $\hat L= \sqrt{\Gamma} (\hat{Q} + i\hat{P})$ describing the coupling to the environment. The first term corresponds to the unitary dynamics given by the Hamiltonian 
\begin{equation}
\label{eq:hamiltonian}
\hat{H} = \frac{1}{2} \hat{P}^2 + \frac{\beta^2}{4}\hat{Q}^4 - \frac{1}{2} \hat{Q}^2 +  \frac{\Gamma}{2}(\hat{Q}\hat{P}+ \hat{P}\hat{Q}) -\frac{g}{\beta}\hat{Q} \cos{(\Omega t)},
\end{equation} 
where $\hat{Q}=\hat x \sqrt{\frac{m \omega_0}{\hbar}} $ and $\hat{P}= \hat p \sqrt{\frac{1}{m \omega_0 \hbar}} $ are dimensionless versions of the position and momentum operators, and $\beta^2 = \frac{\hbar}{ml^2 \omega_0}$ is a dimensionless parameter that defines the scale of the phase space relative to Planck's constant~\cite{Brun:1996,Ota:2005,Kapulkin:2008}. Note that as $\beta \to 0$ the classical limit is achieved.

The final step in the description of our model is to move from the master equation (\ref{eq:ME}), which corresponds to the ensemble averaged evolution of the open quantum system, to an equation that describes a single quantum system being continuously monitored. Such a description is given by quantum trajectories governed by a stochastic Schr\"odinger equation (SSE). Here we will focus on the so-called diffusive trajectories which, for a single noise term, are given in Ito form by~\cite{Rigo:1997,Wiseman:2001b}
\begin{eqnarray}
\label{eq:SSEITO}
\mathrm{d} \ket{\psi} &=& \left(-i \hat H  -\frac{{\hat L}^\dagger \hat L}{2} + \langle {\hat L}^\dagger \rangle \hat L -\frac{ \langle {\hat L}^\dagger \rangle \langle \hat L \rangle}{2}  \right) \ket{\psi} \mathrm{d}t \nonumber \\  &+& \left(\hat L - \langle \hat L \rangle \right)\ket{\psi} \mathrm{d}\xi.
\end{eqnarray}
Here the noise term d$\xi$ is a complex Wiener process with zero mean (E[d$\xi$]$=0$) and correlations given by
\begin{equation}
\label{eq:corr}
\mathrm{d}\xi \, \mathrm{d}\xi^* = \mathrm{dt} \qquad \mathrm{and} \qquad \mathrm{d}\xi \, \mathrm{d}\xi = u\, \mathrm{dt}, 
\end{equation}
where the complex number $u \equiv \vert u\vert e^{-2 i \phi}$ must satisfy the condition $|u| \leq 1$ \cite{Rigo:1997, Wiseman:2001b}. We can then write the complex Wiener process as
\begin{equation}
\mathrm{d}\xi = e^{-i\phi}\left( \sqrt{\frac{1 + |u|}{2}}\, \mathrm{d}W_1 + i \,\sqrt{\frac{1-|u|}{2}} \, \mathrm{d}W_2  \right), 
\label{eq:noise}
\end{equation}
where $\mathrm{d}W_1$ and $\mathrm{d}W_2$ are independent real Wiener processes. Note that the amplitude $|u|$ and phase $\phi$ fully characterise the noise process and therefore different choices of $u$ correspond to particular ways of unraveling the master equation into stochastic trajectories.

\begin{figure}[ht]
\centerline{\includegraphics[width=0.4\linewidth]{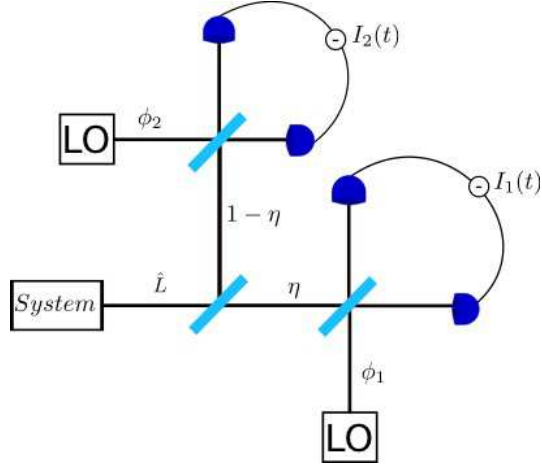}}
\caption{Monitoring scheme for the unraveling parametrisation in terms of $u$. The first beam splitter has transmittance $\eta$ while the ones at the detectors end are balanced. The local oscillators (LO) used in the homodyne-like measurements have phases $\phi_1$ and $\phi_2$.}
\label{fig:full}
\end{figure}

At this point it is important to recognise that $u$, more than providing a convenient mathematical parametrization of the unravelings, also bears a direct connection to a physical way of continuously monitoring the quantum system~\cite{Wiseman:2001b}. For example, if the dissipation operator L describes an optical channel observed using the scheme shown in Fig.~\ref{fig:full}, there is a direct relationship between the beam splitter ratios and phases indicated in the figure and the value of $u$ corresponding to that measurement:
\begin{equation}
u^* = \eta e^{2i \phi_1} + (1-\eta)e^{2i \phi_2}
\end{equation}
and the complex Wiener noise can be written as
\begin{equation}
\mathrm{d}\xi^* = \sqrt{\eta} e^{i\phi_1} \mathrm{d}W_1 + \sqrt{1-\eta} e^{i \phi_2} \, \mathrm{d}W_2.
\label{eq:noise_exp}
\end{equation}
By comparing equation~(\ref{eq:noise}) with equation~(\ref{eq:noise_exp}), we can immediately establish a direct connection between $u$ and the physical parameters $\eta$, $\phi_1$ and $\phi_2$ of the monitoring. Note that these diffusive quantum trajectories correspond to homodyne-like measurements that are routinely implemented in quantum optical setups and that have been measured recently in superconducting qubit systems~\cite{ Murch:2013, Roch:2014, Campagne-Ibarcq:2016}.

Previous works on quantum Lyapunov exponents and the quantum to classical transition have adopted a fixed monitoring strategy (a particular case of equation~(\ref{eq:SSEITO}) for a given choice of $u$ and $\hat L$ operator) corresponding to either a continuous position measurement ($u=1$ and $\hat L =\hat x$)~\cite{Bhattacharya:2000,Habib:2006} or to the quantum state diffusion (QSD) model ($u=0$)~\cite{Ota:2005,Kapulkin:2008}. 

Here, however, we explore different measurement schemes by considering the case where $\vert u \vert=1$ and continuously varying the phase $\phi$. In Fig.~\ref{fig:full} this corresponds to a single homodyne measurement ($\eta =1$) with the local oscillator phase $\phi_1 = \phi$ being varied. From the expression for the measurement signal $I_1 \mathrm{d}t =\sqrt{\eta}\langle e^{-i\phi_1} \hat{L} + e^{i\phi_1}\hat{L}^\dagger \rangle + \mathrm{d}W$~\cite{Wiseman:2001b}, we can see that the choice of local oscillator phase corresponds to the measurement of a particular quadrature, e.g. choosing $\phi = 0$ leads to a measurement of $Q$, while $\phi = \pi/2$ will similarly measure the $P$ component.  Thus we can explore the onset of complexity in quantum systems undergoing monitoring strategies that are routinely implemented in experiments. 

\section*{Results}

We are now in the position to investigate the dynamics of a chaotic quantum Duffing oscillator under continuous monitoring. To establish a quantitative picture of the level of chaos in the dynamics, we calculate the quantum Lyapunov exponent, defined as $\lambda =\lim_{t \to \infty} \lim_{d_0 \to 0} \log{\left(d_t/d_0\right)}/t$, by adapting the usual classical procedure~\cite{Wolf:1985}. Two quantum trajectories, starting from initial coherent states that are displaced from each other by a small distance $d_0$, are evolved stochastically via equation~(\ref{eq:SSEITO}) under the same noise realization which corresponds to the same measurement results. The phase-space distance $d_t=\sqrt{(\Delta x(t))^2 + (\Delta p(t))^2}$ is defined in terms of the expectation values of the position and momentum operators for the two evolving trajectories with $\Delta x(t) = \langle Q_{1} \rangle - \langle Q_{2} \rangle$ and $\Delta p(t) = \langle P_{1} \rangle - \langle P_{2} \rangle$. For the numerical calculations, one of the trajectories is periodically reset towards the other one to remain within the linear regime, and $\log{\left(d_t/d_0\right)}$, calculated before every reset, is averaged over time. Convergence occurs within $500$ cycles of the driving term, and the final Lyapunov exponent is obtained after averaging out over multiple realisations (20 runs) of the stochastic noise. The numerical calculations can be computationally intensive as the size of the system is increased. However, since the state in a chaotic system is confined to the strange attractor, we need only a large enough basis size to encompass this region of phase space for a given choice of $\beta$. In this work we vary the scaling parameter from $\beta = 1$ to $0.1$, which requires a range of basis size from $N= 35$ to $200$ (using the harmonic oscillator energy eigenstates). 

\begin{figure}[ht]
\centerline{\includegraphics[width=0.5\linewidth]{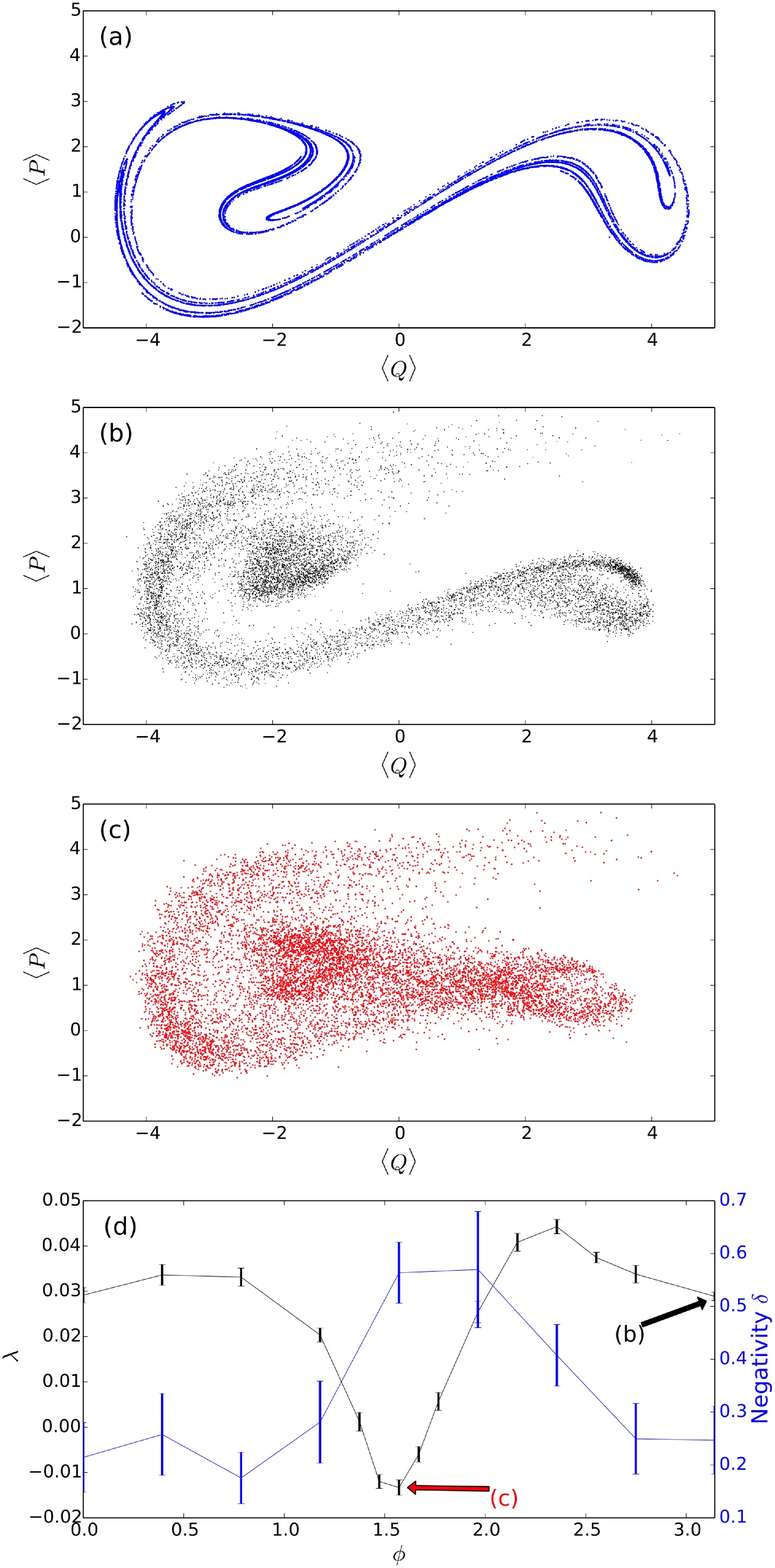}}
\caption{(a) Strange attractor for the classical Duffing oscillator with parameters $\Gamma = 0.10$, $g=0.3$ and $\Omega = 1$. (b) Quantum Poincar\' e section for a single trajectory with 1000 points for $\phi = \pi$ with points taken once every driving period ($t = 2\pi n$). (c) Quantum Poincar\' e section for $\phi = \pi/2$. (d) Average Lyapunov exponent $\lambda$ (black curve) and average negativity $\delta$ (blue curve) as a function of $\phi$ for $|u| = 1$, $\beta = 0.3$, $\Gamma = 0.10$, $g = 0.3$ and $\Omega = 1$. The averages are constructed from the Lyapunov exponents and negativities calculated for 20 different individual trajectories. Error bars are given by the standard error in the mean. The arrows indicate the choice of $\phi$ that correspond to the Poincar\'e sections in (b) and (c). For (b), (c) and (d), with the choice $\beta = 0.3$, a basis size of $N =65$ is used. }
\label{fig:CL}
\end{figure}

The effect of the monitoring angle on the quantum Lyapunov exponents is shown in Fig.~\ref{fig:CL}-d for $\Gamma = 0.10$, $g=0.3$, $\Omega = 1$, and $\beta=0.3$. The quantum dynamics is chaotic ($\lambda>0$) for most choices of the phase $\phi$, with the quantum Poincar\' e section (Fig.~\ref{fig:CL}-b) roughly following the classical strange attractor, which is shown in Fig.~\ref{fig:CL}-a for comparison. For $\phi\approx \pi/2$, however, the quantum attractor is significantly blurred (Fig.~\ref{fig:CL}-c) leading to a strong suppression of chaos. This shows that we can tune the behaviour of the system from chaotic to regular by simply changing which quadrature is measured in the homodyne setup. 

\begin{figure}[ht]
  \centerline{\includegraphics[width=0.5\linewidth]{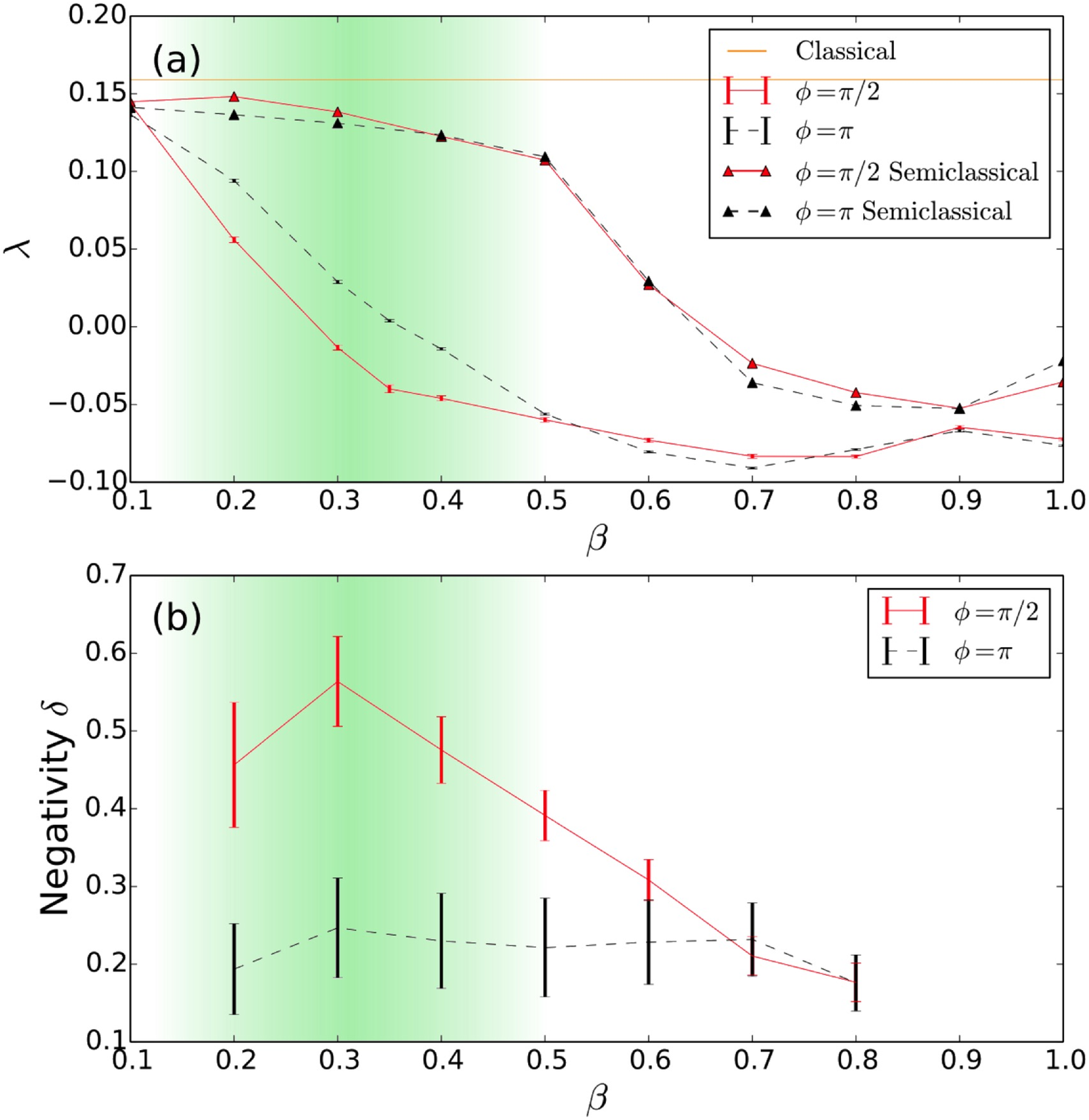}}
   \caption{(a) Average quantum Lyapunov exponent $\lambda$ for $|u| = 1$, $\Gamma = 0.10$, $g=0.3$ and $\Omega = 1$. The points are constructed from averaging 20 different noise realisations, each starting with a pair of coherent states. Error bars give the standard error from the mean. In (a) $\lambda$ is shown as a function of the macroscopicity parameter $\beta$ for the full quantum simulation with two values of the phase $\phi = \pi$ (black, dashed) and $\phi = \pi/2$ (red, solid). The straight line at the top corresponds to the classical Lyapunov exponent $\lambda_{cl}=0.16$ for these parameters. $\lambda$ is also plotted as a function of $\beta$ using a semiclassical approximation (triangles) to obtain the equations of motion, again with the same phases. (b) Negativity $\delta$ of the Wigner function, is averaged over the 20 trajectories for the same parameters and phases as in (a) and plotted as a function of $\beta$. The region where we see a pronounced difference between monitoring strategies is highlighted in green.}
   \label{fig:full2}
\end{figure}

It is evident that in the classical limit, there is no dependence on the monitoring scheme, so there must be a value of $\beta$ beyond which the choice of monitoring can have an effect on the complex behaviour of the system. To investigate that, we plotted in Fig.\ref{fig:full2} the quantum Lyapunov exponents for $\phi=\pi$ (dashed black) and $\phi=\pi/2$ (solid red), corresponding approximately to the maximum and minimum values of $\lambda$, as a function of our macroscopicity parameter $\beta$. The curves show that 
for large $\beta$, the quantum Lyapunov exponent is always negative and it is not significantly affected by the choice of $\phi$. This is the region where the quantum noise is dominant, chaos is suppressed, and the exact form of the monitoring is irrelevant. In the opposite limit, the quantum curve is always positive and should approach the classical value of $\lambda_{cl}=0.16$ for small enough $\beta$. This region also shows very little dependence with $\phi$, but now for a different reason: in this limit the classical dynamics prevails over the quantum noise and the choice of measurement ceases to affect the system. However, there is an intermediate region, highlighted in Fig.~\ref{fig:full2}, where there is a noticeable dependence on $\phi$. This is exactly the window of the macroscopicity parameter where controlling the onset of chaos through quantum measurements is possible. 

To utilise this window of control and predict the monitoring parameters that provide minimum or maximum Lyapunov exponents for a given chaotic system, one first needs to understand the physical mechanism behind this dependency with the angle $\phi$. A first hint towards the explanation comes from the semiclassical results shown by the top two curves in Fig.~\ref{fig:full2}, obtained using a Gaussian approximation~\cite{Pattanayak:1994}. We see a negligible difference between the Lyapunov exponents for the two curves, indicating that the difference we see in the quantum dynamics cannot be explained by the Gaussian approximation. This approximation retains the measurement terms and the stochastic aspect of the dynamics, but restricts the state to remain as a Gaussian in phase space. The latter aspect prevents the formation of the complex interference fringes that we see in the full quantum evolution (see Fig.~\ref{fig:neg} and Supplemental Material for an animation of the dynamics for $\phi=\pi/2$ (see Supplementary video S1 online) $\phi=\pi$ (see Supplementary video S2 online) ) and indicates that the effect we observed is intrinsically quantum, arising from the interplay between the interference generated by the nonlinear dynamics and the way different monitoring strategies destroy them. 

\begin{figure}[ht]
  \centerline{\includegraphics[width=0.5\linewidth]{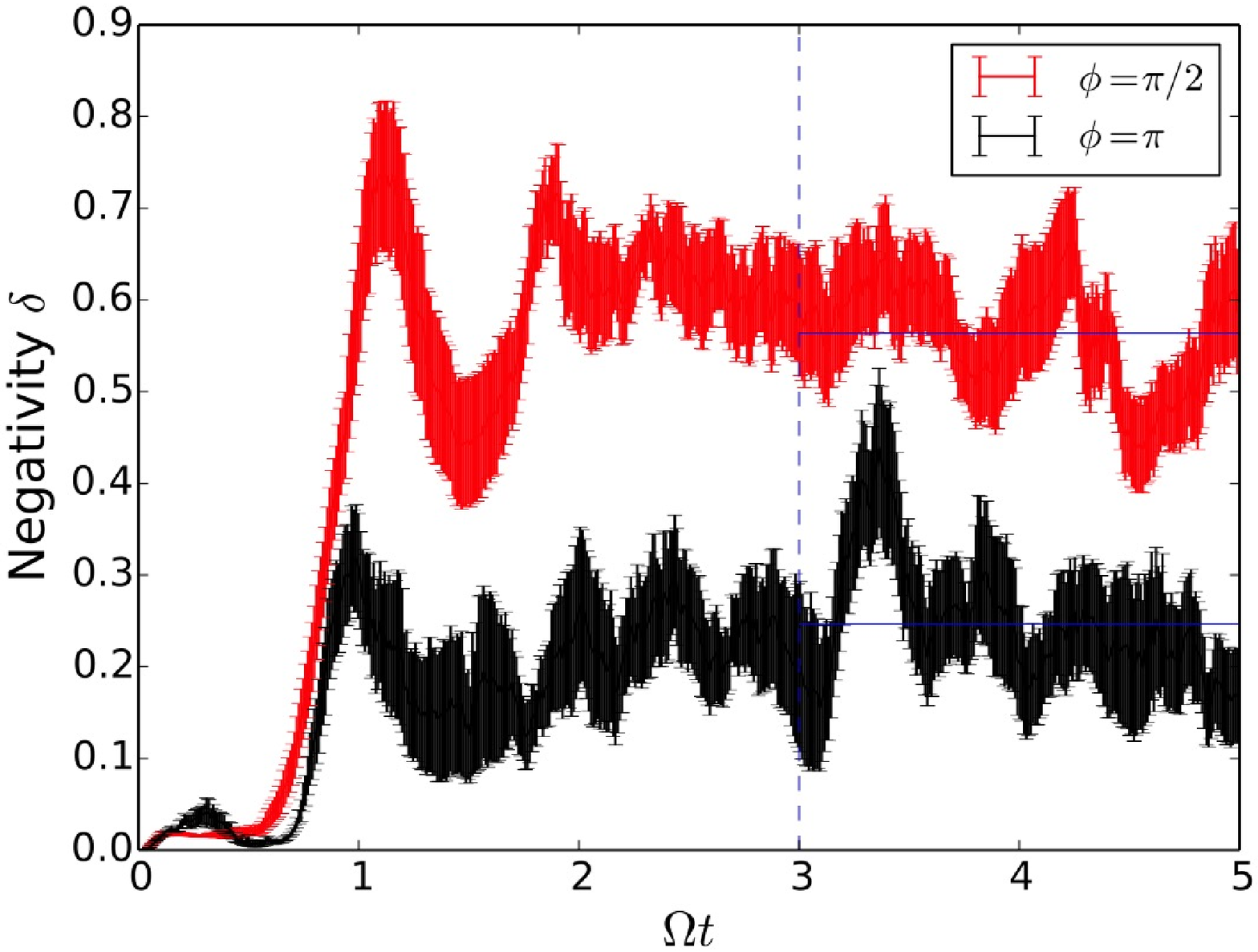}}
   \caption{Evolution of the negativity of the Wigner function for $\phi=\pi/2$ (red) and $\phi=\pi$ (black) averaged over $20$ noise realisations. The horizontal lines show the average negativity for the last $2$ forcing periods, a value that is used in Fig.~\ref{fig:CL}-d. We see that smaller (higher) values of negativity correspond to higher (smaller) values of the Lyapunov exponents.   
   }
   \label{fig:negtime}
\end{figure}

In order to assert that interference effects are indeed the key factor at play, we must quantify the level of interference present in the evolution. To do this we use the negativity of the Wigner function which has previously been proposed as an indicator of non-classicality~\cite{Kenfack:2004a} and is defined as $\delta_\psi = \int \int |W_\psi(q,p)|~\mathrm{d}q~\mathrm{d}p - 1$. Here, both the Wigner function and the negativity are calculated for the pure state $\psi$ evolved in each individual quantum trajectory. The average negativity is then found by averaging over the 20 noise realisations $\delta = \sum_i^{M=20} \delta_{\psi_i}/M$. In Fig.~\ref{fig:negtime}, where we show the average negativity as a function of time, we see that for both choices of monitoring angles in the figure (red curve for $\phi=\pi/2$ and black for $\phi=\pi$) the negativity starts from zero (initial coherent state) and has a surge at around $0.8 \Omega t$. This is the time that it takes for the quantum state to start probing the skeleton of the classical attractor and start developing fringes in the Wigner function (see Supplemental Material for an animation of the dynamics for $\phi=\pi/2$ (see Supplementary video S1 online) and  $\phi=\pi$ (see Supplementary video S2 online)). After this build up period, the effect of the monitoring on the dynamics becomes evident: for $\phi=\pi/2$ the negativity fluctuates around higher values than for $\phi=\pi$. To make a connection with the Lyapunov exponents calculated previously, we averaged the negativity from Fig.~\ref{fig:negtime} for the last 2 forcing periods and plotted the results in Fig.~\ref{fig:CL}-d. What we see is a clear anti-correlation between the negativity and the quantum Lyapunov exponents, which is also observed for the different values of $\beta$ in Fig.~\ref{fig:full2}. The large negativity at $\phi\approx \pi/2$ explains the dip in the Lyapunov exponent: the larger the interference effects, the further the quantum system is from the classical behaviour, leading to a stronger suppression of chaos. 

The remaining issue to be explained is why the suppression is stronger at that particular measurement angle. The best way to understand the role of the phase $\phi$ is to examine a simple class of states that present interference. Here we look at superpositions of coherent states in the form $|\psi_0\rangle = c_+|\alpha\rangle + c_- |-\alpha\rangle$ (Schr\"odinger cat state), with $\alpha=\vert \alpha \vert e^{i \varphi}$. The interference fringes in these states have a well defined structure, being aligned along the direction defined by the angle $\varphi$. The evolution of the state conditioned on the measurement only is given by equation~(\ref{eq:SSEITO}) with $H=0$. Looking just at the noise term $\left(\hat a - \langle \hat a \rangle \right)\ket{\psi} \mathrm{d}\xi$ for $|u| = 1$, we have
\begin{equation}
\mathrm{d}|\psi_0\rangle =\vert \alpha \vert \left(c_+ e^{i\left(\varphi -\phi\right) }|\alpha\rangle - c_- e^{i\left(\varphi -\phi\right) } |-\alpha\rangle  \right) \mathrm{d}W,
\label{eq:CATevo}
\end{equation}
where we have assumed that the initial coefficients are equal ($c_+=c_-$) and that $\alpha$ is large, such that $\langle -\alpha \vert \alpha \rangle \approx 0$.  

For each term the evolution is given by
\begin{equation}
\mathrm{d}c_+ =  c_+ \vert \alpha \vert e^{i\left(\varphi -\phi\right) } \mathrm{d}W,
\label{eq:CATevopos}
\end{equation}
\begin{equation}
\mathrm{d}c_- =  -c_- \vert \alpha \vert e^{i\left(\varphi -\phi\right) } \mathrm{d}W.
\label{eq:CATevoneg}
\end{equation}

From these equations one can see that when the monitoring angle $\phi$ is parallel to the interference fringes ($\varphi-\phi =0$), then in a short time the system stochastically evolves to one of the components of the original superposition and interference fringes quickly disappear (see Supplemental Material for an animation of this effect on a cat state for $\phi=\pi$ (see Supplementary video S3 online) and $\phi=\pi/2$ (see Supplementary video S4 online)). On the other hand, when the measurement direction is perpendicular to the fringes ($\varphi-\phi = \pi/2$), the short term evolution corresponds to a phase rotation between the two components and the fringes survive for longer. Therefore, the efficacy of the the stochastic term in equation~(\ref{eq:SSEITO}) in eliminating interference depends directly on the alignment between $\phi$ and $\varphi$. 
\begin{figure*}[ht]
  \centerline{\includegraphics[width=\linewidth]{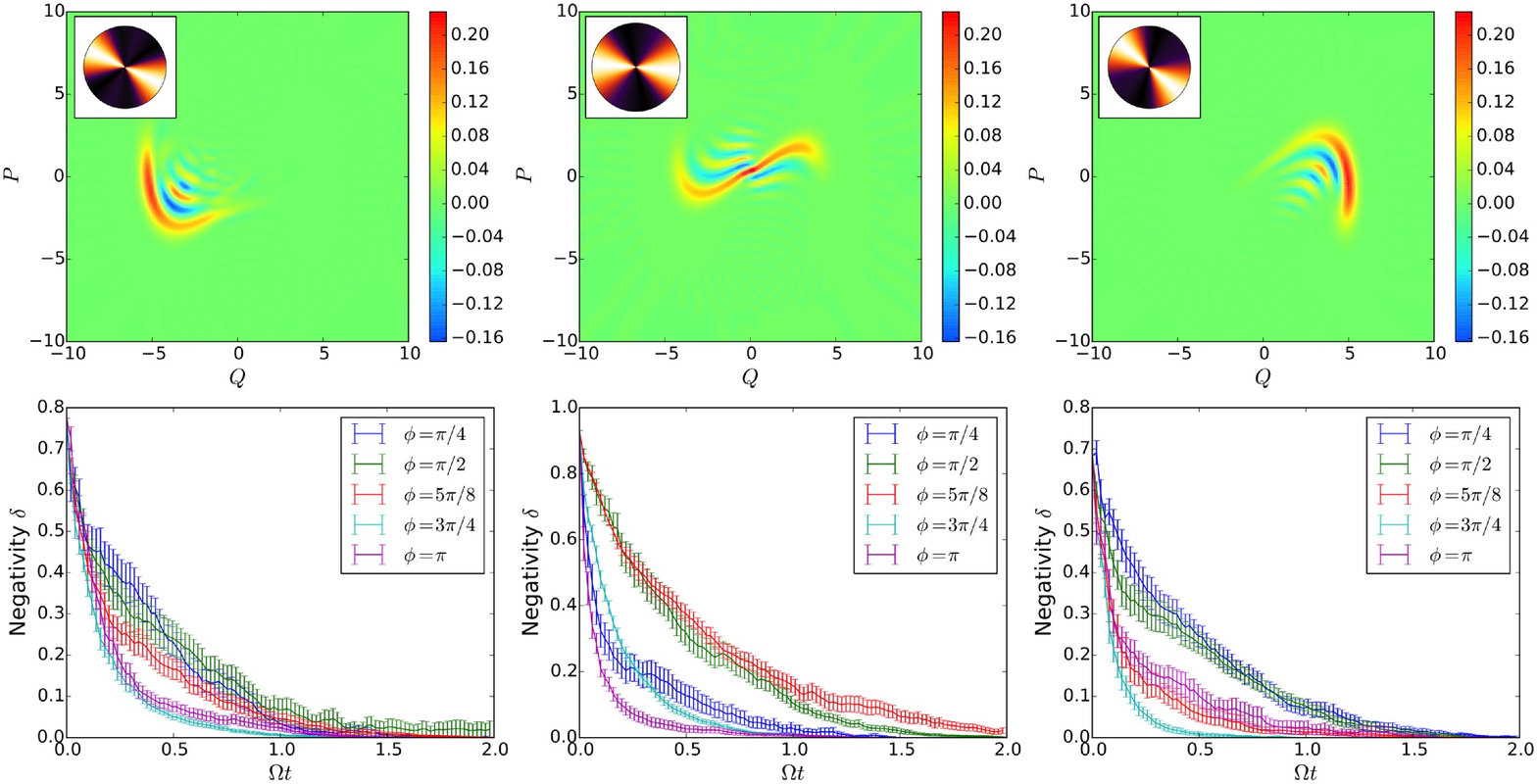}}
   \caption{(Top) Snapshots of the Wigner function (for a single quantum trajectory) for the full dynamics at three different times (for $\Gamma = 0.10$ and $\beta = 0.3$). These states are evolved solely under the monitoring dynamics and the decay of the negativity is shown in the bottom plots, averaged over 20 noise realisations. As in the case of cat states, negativity decays faster when the monitoring angle is parallel to the interference pattern. This is confirmed by the insets in the top panels where the density plots of the negativity decay rates, obtained by fitting the curves in the bottom plots, are shown with the minimum (maximum) decay rates corresponding to the darker (brighter) colors.      
   }
   \label{fig:neg}
\end{figure*}

This is exactly what explains the dependency of the quantum Lyapunov exponent with the monitoring parameter. Given the complexity of the dynamics and the geometric structure of the strange attractor in phase space, it is non-trivial to justify that there is a privileged direction where this effect can take place. However, by following the dynamics of the Wigner function for a single trajectory in real time (see Supplemental Material for an animation of the dynamics for $\phi=\pi/2$ (see Supplementary video S1 online) and $\phi=\pi$ (see Supplementary video S2 online)), it is possible to distinguish certain structures that repeat over time. These structures, representing the stretching region around the origin and also the left and right bending regions of the classical strange attractor, are depicted in the snapshots of the Wigner function of Fig.~\ref{fig:neg}. Even though the interference in these plots are not perfectly aligned, they are concentrated in the range of angles orthogonal to the ones that lead to higher negativity (around $\phi= \pi/2$). This dependency with the angle is quantified by calculating the decay rate of the negativity for the evolution of these states under monitoring dynamics only (bottom plots in Fig.~\ref{fig:neg}), and they match the region where the Lyapunov exponent dips in Fig.~\ref{fig:CL}-d.  

\begin{figure}[ht]
  \centerline{\includegraphics[width=0.5\linewidth]{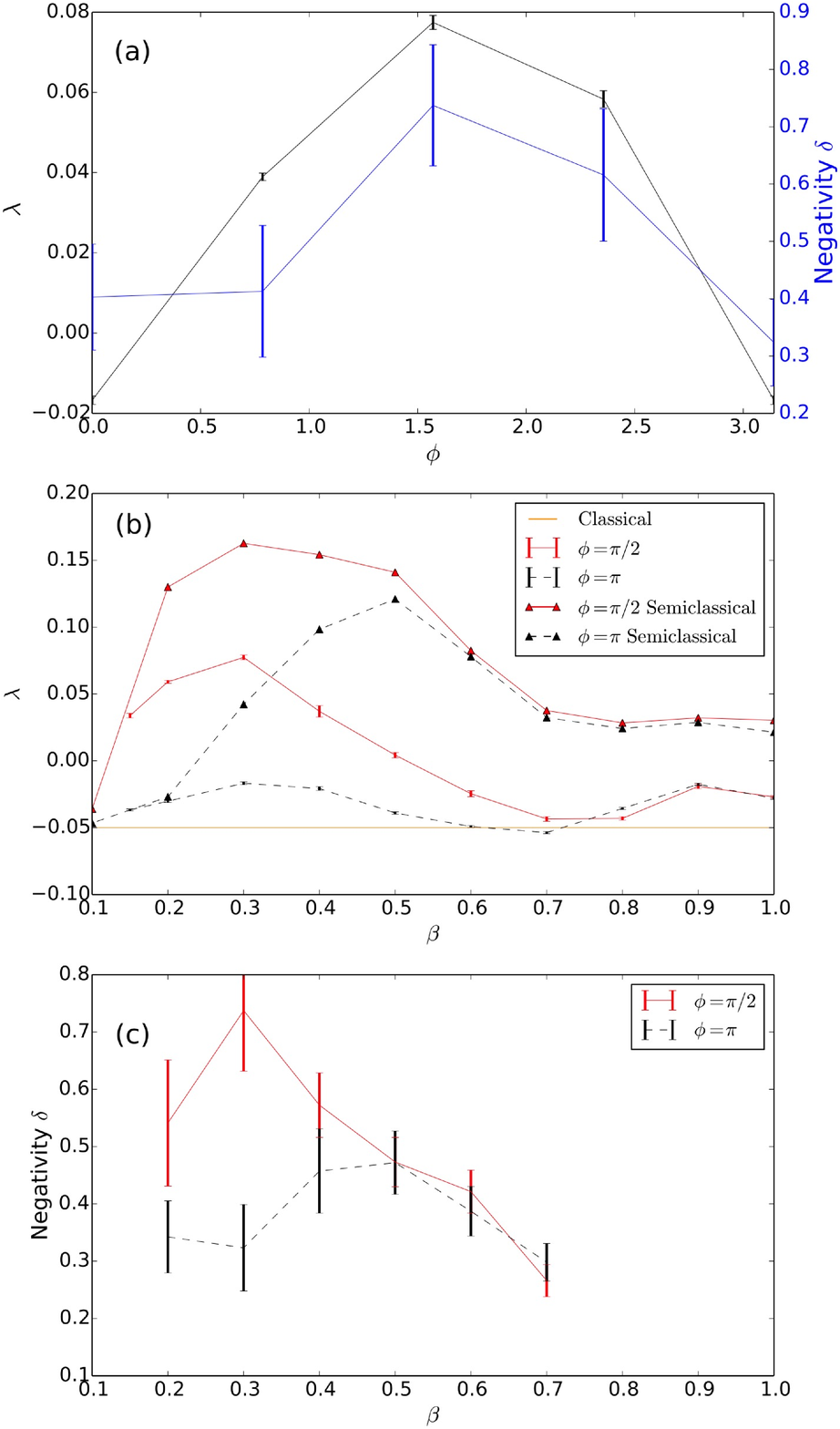}}
   \caption{(a) Lyapunov exponent (black) and average negativity (blue) as a function of the monitoring parameter $\phi$ for $\Gamma = 0.05$. The Lyapunov exponents for $\phi = \pi$ (black, dashed) and $\phi = \pi/2$ (red, solid) are given in (b) as a function of the macroscopicity parameter $\beta$ for the quantum and semiclassical (triangles) for values from $0.1$ to $1.0$. The points here are constructed from 10 different noise realisations. The classical Lyapunov exponent in this case is $\lambda_{cl} = -0.05$ and is represented by the horizontal line in the plot. (c) Average negativity of the Wigner function as a function of $\beta$ for the same phases as in (b) . Each point is constructed from 20 different noise realisations. The error bars give the standard error in the mean.       
   }
 \label{fig:regular}
\end{figure}

Given the evidence presented so far, it is tempting to always associate the presence of negativity in the Wigner function with suppression of chaos. However, in certain cases a higher value of negativity seems to be connected with enhancement of chaos, as shown in Fig.~\ref{fig:regular}, where the quantum Lyapunov exponents for $\Gamma=0.05$ are shown (all other parameters are as in Fig.~\ref{fig:CL}) . This is an interesting case recently investigated by Pokharel et al.~\cite{Pokharel:2016} where the classical dynamics is regular but chaos can emerge quantum mechanically. While this seems to be a counter example of our discussion so far, the fact that the dependency of the Lyapunov exponent with the monitoring angle is also seen in the semiclassical calculations (see Fig.~\ref{fig:regular}-b), indicates that a different mechanism is at play in this case, overshadowing the effects of negativity. 

\begin{figure}[ht]
  \centerline{\includegraphics[width=\linewidth]{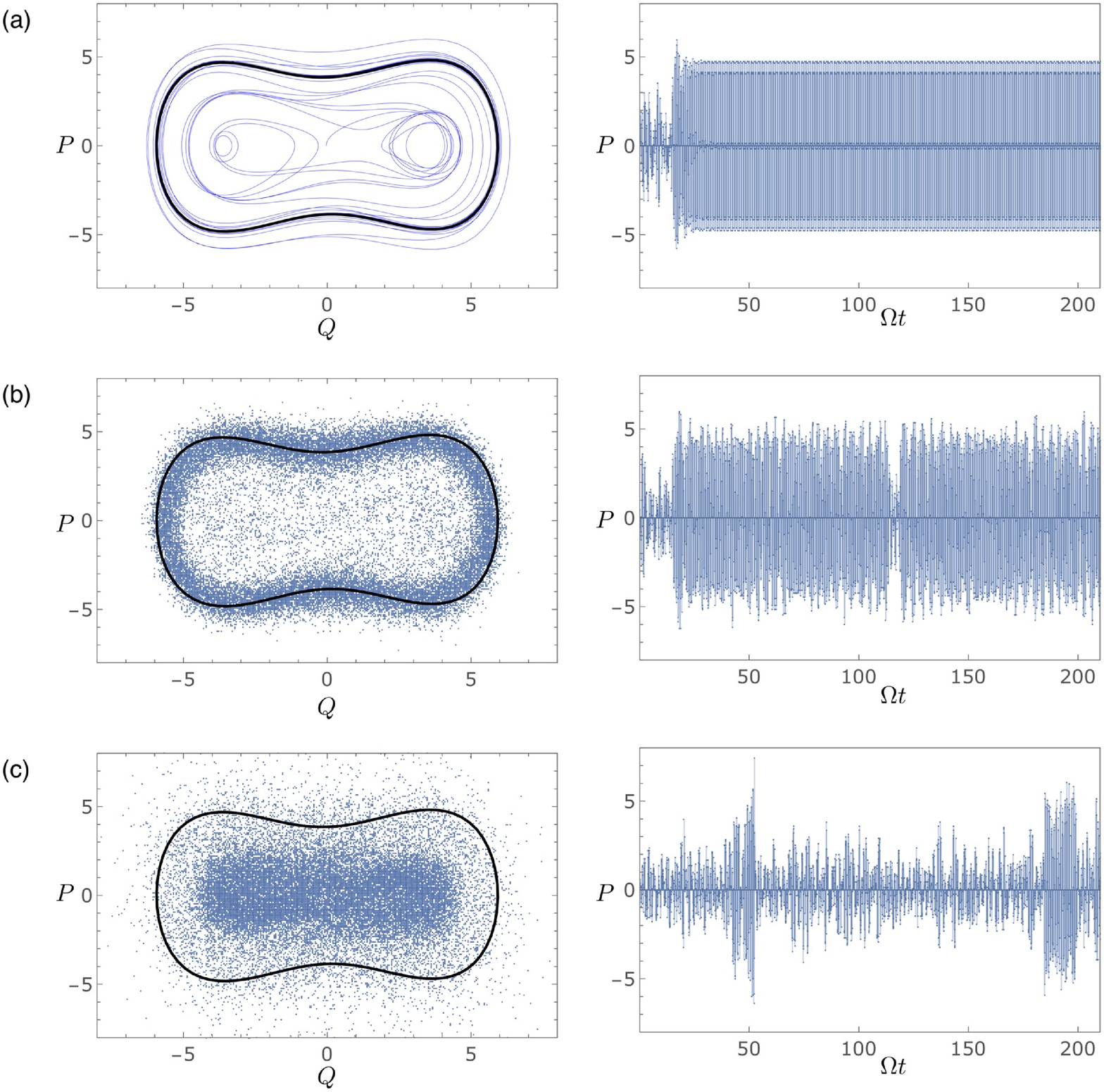}}
   \caption{Phase space projection (left column) and momentum time series (right column) for: (a) classical; and semiclassical with (b) $\phi=\pi$ and (c) $\phi=\pi/2$. The bold black line corresponds to the classical dynamics with the transient removed. Classically, the central chaotic region is only visited in the initial transient before the system settles in its periodic behaviour. Semi-classically, the system can transition from one region to another. Different choices of the monitoring parameter $\phi$ induce preferences towards the classical periodic orbit (b), or the irregular region (c).}
   \label{fig:regular2}
\end{figure}

Indeed, simulations of the dynamics using the Gaussian approximation show that for the phase $\phi$ corresponding to the smallest value of the Lyapunov exponent, the semiclassical system remains most of the time concentrated along the classical stable orbit, rarely making incursions into the central region corresponding to the classical chaotic transient (see Fig.~\ref{fig:regular2}). In the stable region, the quantum state remains mostly Gaussian, explaining the small values for the negativity. On the other hand, for the phase linked to the maximum Lyapunov exponent, the semiclassical system spends more time in the chaotic region, visiting the stable classical orbit from time to time, but eventually coming back. While visiting the chaotic region, the quantum state is allowed to stretch along the unstable direction and then fold, interfering with itself and producing negative values of the Wigner function (see Supplemental Material for animations of the Wigner function for minimum (see Supplementary video S5 online) and maximal (see Supplementary video S6 online) Lyapunov exponents). The existence of negative values is therefore a consequence of a semiclassical dynamical effect of the monitoring process, which, for certain values of $\phi$, induces transitions between the coexisting regular and chaotic regions. These transitions seems to be related to the recent analysis of the quantum-classical correspondence in terms of transient chaos done by Wang et al.~\cite{Wang:2016}. Note, however, that here the noise strength is fixed, it is therefore the form of the coupling between the noise and the system variables, determined by the measurement choice, that dictates the average time spent in each region. Once again, just a change in our measurement parameter allows us to radically alter the complexity of the dynamical evolution of the system. 

In conclusion, our results show that the choice of monitoring plays a crucial role in the emergence of chaos in quantum systems, adding yet another layer of complexity to the already intriguing problem of the quantum to classical transition. We showed that the effect of the measurement choice on the quantum Lyapunov exponent manifests in two distinct ways: at the semiclassical level, by inducing transitions between regions corresponding to a classical periodic orbit and a transient chaotic regime; at the quantum level, by influencing the way interference fringes in the Wigner function are destroyed. In both cases, the more quantumness in the system, as measured by the amount of negativity, the more its dynamical behaviour departs from the classical: by suppressing chaos in the latter and creating it in the former. In the case where the corresponding classical system is chaotic, the effectiveness of certain monitoring schemes in suppressing interference depends on the relative angle between the measurement direction and the fringes induced by the nonlinear dynamics. In this way, we have predictive power over the monitoring parameters that will lead to minimum or maximum quantum Lyapunov exponents by analysing the geometrical structure of the classical attractor. In both cases it is remarkable that despite the fact that the form and amount of dissipation in the system, as well as the system size, are kept constant, we are still able to manipulate the onset of complex behaviour in the system by tuning a purely quantum parameter associated with the appropriately chosen measurement scenario.  

\bibliography{allbibcopy}

\section*{Acknowledgements}

We are grateful to A. Pattanayak, M. Misplon and W. Lynn for enlightening discussions on their recent results~\cite{Pokharel:2016}. This work was supported by the Australian Research Council Centre of Excellence for Quantum Computation and Communication Technology (Project number CE110001027) and the Australian Research Council Future Fellowship (FT120100291).

\section*{Author contributions statement}

J.K.E, J.J.H and A.R.R.C conceived the research, J.K.E conducted the numerical simulations, J.K.E and A.R.R.C wrote the manuscript. All authors contributed to discussions of the simulations and results, and assisted in editing the manuscript.

\section*{Additional information}

\textbf{Supplementary information} accompanies this paper at http://www.nature.com/srep

 \textbf{Competing financial interests:} The authors declare no competing financial interests.

\end{document}